\documentclass[aps,prb,floatfix,twocolumn,footinbib]{revtex4-1} 
\usepackage{epsfig}
\usepackage{hyperref}
\usepackage{cancel} 
\usepackage{amsmath, amssymb} 
\usepackage{graphicx}
\usepackage{epsf}
\usepackage{braket}
\usepackage{appendix}
\usepackage{color}
\usepackage{gensymb}

\graphicspath{{Figures/}}

\begin{document}

\title{Mode selection in InAs quantum dot microdisk lasers using \\ focused ion beam technique}

\author{A.\,A.~Bogdanov$^{1,2,3}$}
\author{I.\,S.~Mukhin$^{1,4}$}
\author{N.\,V.~Kryzhanovskaya$^{1,3}$}
\author{M.\,V.~Maximov$^{1,2,3}$}
\author{Z.\,F.~Sadrieva$^{1,4}$}
\author{M.\,M.~Kulagina$^{2}$}
\author{Yu.\,M.~Zadiranov$^{2}$}
\author{A.\,A.~Lipovskii$^{1,3}$}
\author{E.\,I.~Moiseev$^{1}$}
\author{Yu.\,V.~Kudashova$^{1}$}
\author{A.\,E.~Zhukov$^{1,3,5}$}

\affiliation{$^{1}$St. Petersburg Academic University, 194021 St. Petersburg,  Russia}
\affiliation{$^{2}$Ioffe Institute, 194021 St.~Petersburg,  Russia}
\affiliation{$^{3}$Peter the Great St. Petersburg Polytechnic University, St. Petersburg, Russia}
\affiliation{$^{4}$ITMO University, St. Petersburg, Russia}
\affiliation{$^{5}$St. Petersburg Scientific Center RAS,  St. Petersburg, Russia}
\date{\today}

\begin{abstract}

Optically pumped InAs quantum dot microdisk lasers with grooves etched on their surface by a focused ion beam is studied. It is shown that the radial grooves, depending on their length, suppress the lasing of specific radial modes of the microdisk. Total suppression of all radial modes except for the fundamental radial one is also demonstrated. The comparison of laser spectra measured at 78~K before and after ion beam etching for microdisk of 8~$\mu$m in diameter shows a six-fold increase of mode spacing, from 2.5~nm to 15.5~nm, without significant decrease of the dominant mode quality factor. Numerical simulations are in good agreement with experimental results.
\end{abstract}


\maketitle

Microcavities with the axial symmetry (rings, disks, tori etc) can support so-called whispering gallery modes (WGMs)~\cite{Righini2011,Nosich2008} formed by the total reflection of the electromagnetic wave from the side wall of the cavity. The main advantages of WGM cavities are (i) their small size in comparison with Fabry-Perot resonators~\cite{vahala2003optical}, (ii) high quality factor allowing to accumulate sufficient electromagnetic energy density even for a considerable manifestation of non-linear effects~\cite{braginsky1989quality,matsko2005review,ilchenko2004nonlinear}, and (iii) a narrow resonance line width which allows detection of small spectral resonance shifts induced by coupling with single molecules or nanoparticles~\cite{he2011detecting}. Compact size and high quality factor make lasers based on WGM cavities prospective for telecommunication and optical processing~\cite{hill2004fast,lissillour2001whispering}. 

Using self-organized arrays of quantum dots (QDs) as active region of microdisk and microring lasers provides a considerable decreasing of their threshold current~\cite{ide2005room,kryzhanovskaya2014whispering,mccall1992whispering}. Large localization energy of charge carriers and lateral spacing between QDs results in a drastic reduction of the lateral carrier transport. As a result, an essential decrease in internal Joule losses and non-radiative recombination of carriers on side walls is achieved. 

Distinct feature of a QD active region is a broad gain spectral range arising from the inhomogeneous broadening of the QD luminescence spectrum (typically, about 40~nm for the ground state optical transition)~\cite{nevsky2008narrow,varangis2000low}. If the mode spacing  in a WGM laser is below the gain spectrum width, several laser modes can be excited. A natural way to achieve a single-mode emission is to increase the mode spacing, e.g. through a decrease of the cavity size. However, as was shown in Ref.~\onlinecite{michler2000laser}, laser generation even in small microdisks is unstable regarding the switching of the lasing between neighbour modes. This can occur even under a small variation of pump intensity or external temperature. Moreover, radiation loss exponentially increases with decreasing the cavity size, that results in a drastic growth of the threshold current in ultrasmall WGM lasers~\cite{Righini2011,boriskina2006directional}.

An alternative approach to increase the mode spacing for providing a single-mode emission regime is a selective suppression of undesirable modes by intentionally introduced defects. These defects serve as scatter centers increasing the radiation loss and, hence, the lasing threshold for specific modes. Every mode has unique spatial distribution of electromagnetic field defining the interaction strength between the mode and the defect. By placing the defects in preassigned positions, it is possible to suppress some modes keeping the threshold characteristics of others without essential changes. A role of the defect can be played by, for example, a groove, notch or pit etched on the surface of the cavity~\cite{Backes1998}, a periodical roughness on the sidewall or cavity face~\cite{Elliot1955,urbonas2015ultra}, pedestal of the cavity~\cite{mao2011room}, a particle placed on the cavity face (i.e. an optical antenna)~\cite{gotzinger2001towards}, and so on. An effect of the etched grooves and pits on the laser spectrum of microdisks was shown by us earlier~\cite{Kryzhanovskaya2014}. Mode selection and single-mode emission were achieved by introducing subwavelength sized notches of about 50~nm width and 500~nm depth to the sidewalls of ring shaped QD microcavities with a diameter of 80~$\mu$m~\cite{Schlehahn2013}. Single mode lasing was obtained in a microgear laser (having a periodic roughness on its sidewall) with diameter varying from 2.3 to 3.2~$\mu$m~\cite{Fujita2002}. 

\begin{figure}[htbp]
   \centering
   \includegraphics[width=0.9\linewidth]{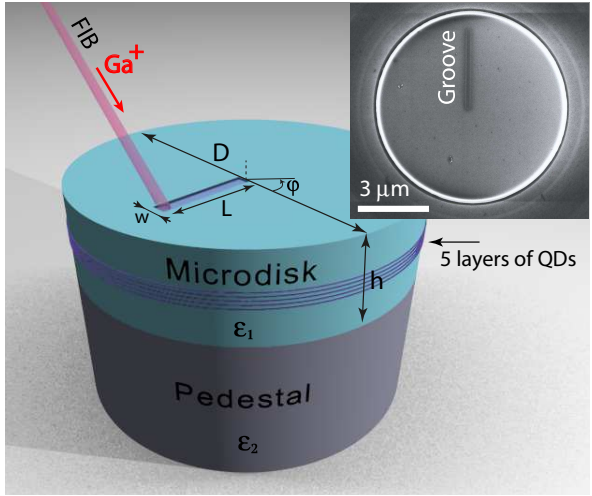} 
   \caption{Schematic representation of GaAs microdisk laser with InAs QD active region on oxidized AlGaAs pedestal. The groove defect on the top face of the cavity is etched by FIB of Ga$^+$. Inset shows SEM image of the top view of the sample -- 8~$\mu$m microdisk with a groove defect. 
   }
   \label{fig:FIB}
\end{figure}

In this Letter we developed and verified a method of controllable suppression of unwanted modes by defects etched on the microdisk top surface using a focused ion beam (FIB) of Ga ions. We have analyzed effect of the position and size of the groove defect on the laser spectrum, threshold power, and quality factor of the modes. Total suppression of all modes except the fundamental radial one is achieved. Increase of mode spacing from 2.5 to 15.5~nm is demonstrated for a QD microdisk laser with a diameter of 8~$\mu$m. 

The samples were grown by solid source molecular beam epitaxy (MBE) on semi-insulating GaAs (100) substrates using a Riber 49 MBE machine. Five layers of InAs/InGaAs QDs were grown in the middle of a 220~nm-thick GaAs waveguide layer placed on 400~nm-thick layer of AlGaAs. Gain spectrum maximum of QD array at 78~K corresponds to the wavelength 1.2~$\mu$m and has width about 35~nm. The cavities with diameter 8~$\mu$m were fabricated using photolithography and ion-beam etching with Ar ions. Next, the AlGaAs pedestal was oxidized into (AlGa)$_x$O$_y$ using the selective oxidation method to provide optical confinement from the substrate side. Groove defects were etched by a focused ion beam of Ga using a Carl Zeiss Neon 40 CrossBeam workstation. Acceleration voltage and current in the Ga$^+$ ion beam were 30~kV and 5~pA, respectively. Sketch of the sample is shown in Fig.~\ref{fig:FIB}.

Optical pumping was provided by a continuous wave Nd:YAG laser emitting at the second harmonic ($\lambda = 532$~nm) with power varying up to 200~mW. The pump laser beam was focused by an Olympus LMPlan IR 100 objective with NA~=~0.8. Photoluminescence (PL) radiation was collected by the same lens and analyzed by a FHR1000 monochromator equipped with a Horiba Symphony cooled multichannel InGaAs photodetector with a resolution of 0.03~nm. 

\begin{figure}[htbp]
   \centering
   \includegraphics[width=1\linewidth]{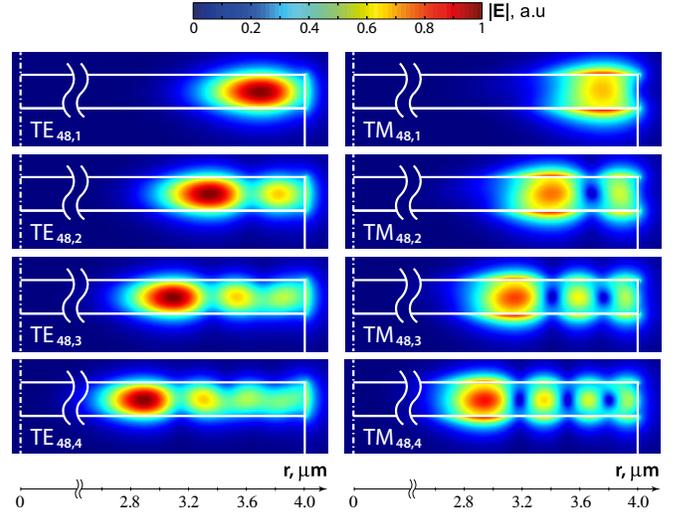} 
   \caption{Distribution of electric field amplitude $|\mathbf E|$ for fundamental ($q$~=~1) quasi-TE and quasi-TM WGMs with azimuthal number $m$~=~48 in microdisk cavity of 8~$\mu$m in diameter. Polarization and mode indices are shown in the panels. Dielectric function of the cavity and pedestal are taken equal to 11.4 and 2.5, respectively. The hight of the cavity is $h$~=~220~nm.}
   \label{fig:field_distribution}
\end{figure}

Finding wavelength and electric field distribution of modes in a resonator of an arbitrary shape requires a 3D numerical simulation. Fortunately, rotational symmetry of WGM resonators allows to reduce this problem from 3D to 2D case owing to known azimuthal dependence given by $\exp(\pm i m\varphi)$. Index $m$ is an azimuthal mode number and $\varphi$ is the azimuthal angle. The resulting 2D problem can be solved numerically or using analytical approximations~\cite{matsko2006optical}. Here, we analyze it numerically with COMSOL Multiphysics software using the approach developed in Ref.~\onlinecite{oxborrow2007traceable}. In the simulation we take into account the (AlGa)$_x$O$_y$ pedestal with a height of 400~nm (as homogeneous material with permittivity 2.5), GaAs microdisk cavity with a height of 220~nm (as homogeneous material with permittivity 11.4), and surrounding air. The diameter of the microcavity and pedestal are supposed to be the same and equal to 8~$\mu$m.

Preferably, WGM modes are characterized by radial ($q$) and axial ($p$) mode numbers having clear physical meaning since it is number of maxima in the corresponding direction. In general case, polarization of eigenmodes in a cylindrical cavity is hybrid one. However, in the case of a weak hybridization, when the modes are nearly TE or TM polarized, we will use the prefix "quasi-". Quasi-TM modes have weaker optical confinement, especially in thin microdisks. If a microdisk is thinner than $\lambda/(2n)$, quasi-TM modes can be neglected (see Ref.~\onlinecite{Ho1991}). However in our case, the microdisk thickness is about $0.6\lambda/(n)$, and both quasi-TE and quasi-TM modes are present  in the luminescence spectrum. 

Several quasi-TE and quasi-TM modes with $p = 1$, $m = 48$ and different radial indices $q$ are shown in Fig.~\ref{fig:field_distribution}. The asymmetry of the field distribution is explained by the different dielectric contrasts on the top and bottom interfaces of the microdisk. One can see that the maximum of electric field distribution shifts to the center of the microdisk with increasing of the radial number $q$. It follows from the results presented that etched on the top surface and starting from the microdisk center radial groove of 3.2~$\mu$m or shorter length should not affect the modes with radial number $q = 1$. At the same time, all the modes with $q > 1$ should scatter on the defect and could have considerable loss suppressing laser generation at these modes. The groove with the length exceeding 3.2~$\mu$m should result in increasing of scattering loss for the modes of first and second radial orders.

\begin{figure}[htbp]
   \centering
  \includegraphics[width=1\linewidth]{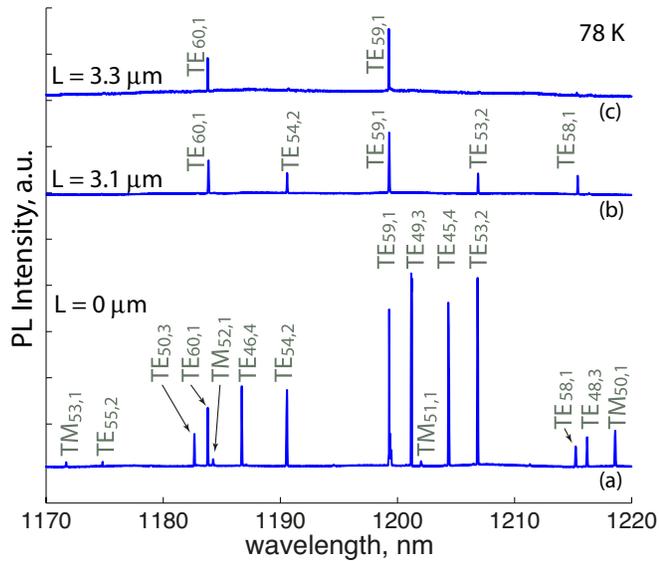} 
   \caption{PL spectrum of microdisk lasers without groove defect (a), with groove of 3.1 (b) and 3.3~$\mu$m (c) in length at temperature 78~K.}
   \label{fig:PL_spectra}
\end{figure}

To confirm this hypothesis a radial groove defect of 10~nm deep, 200~nm wide and 3.1~$\mu$m (3.3~$\mu$m for other sample) long was FIB-etched. Scanning electron microscope (SEM) image of the sample is shown in the inset of Fig.~\ref{fig:FIB}.

PL spectra of the samples without a groove and with 3.1- and 3.3~$\mu$m-long grooves are shown in Fig.~\ref{fig:PL_spectra}. The spectra are shifted along the ordinate axis for clearness. Power of optical pumping was 1.3~mW that exceeded the lasing threshold, which was about several hundreds of $\mu$W for all the samples. The sharp peaks in the spectra correspond to quasi-TE and quasi-TM modes of different radial and axial orders. Identification of the modes was done using numerical simulation. One can see that the modes of radial order $q = 1, 2, 3, 4$ are observed in the microdisk without the groove defect (see Fig.~\ref{fig:PL_spectra}a). The average spacing between adjacent quasi-TE modes is about 2.5~nm. At the same time, spacing between modes with the same azimuthal order and neighbouring radial one is much bigger, about 15.5~nm.

Etching of the 3.1~$\mu$m long groove defect results in the suppression of the modes with radial numbers $q = 3, 4$ and significant thinning of the spectrum (see Fig.~\ref{fig:PL_spectra}b). Further increasing of the groove length up to 3.3 ~$\mu$m results in suppression of the modes with radial number $q = 2$ (see Fig.~\ref{fig:PL_spectra}c). Thus, only two quasi-TE modes ($q = 1$ with $m = 59$ and 60) remain in the spectrum.

A negative effect of the groove defect is an increase of the threshold power. As the length of  10-nm deep defects changes from 0 to 3.3~$\mu$m, threshold power grows from 0.3 to 0.6~mW (see Table~\ref{tab:threshold}). The increasing of the threshold power is a result of quality factor decrease which can be explained by an enhancement of scattering loss and non-radiative recombination of photoexcited carriers at structure defects arising under FIB action. Measurement of the dominant mode quality factor, defined as $\lambda/\Delta\lambda$, in the vicinity of the lasing threshold yields $3.0\times10^4$ for the initial microdisk and $2.0\times10^4$ after the groove etching.   

\begin{table}[htbp]
\centering
\caption{\bf Dependence of threshold power on groove geometry for TE$_{59,1}$ mode at 78~K}
\label{tab:threshold}
\begin{tabular}{ccccc}
\hline
 Groove  & Groove & Threshold &$\Delta\lambda$ & Quality  \\
 length, $\mu m$ & depth, $nm$ & power, $mW$ & $pm$ & factor\\
\hline
0  & 0 & 0.3 & 40 & 3.0$\times$10$^4$\\
3.3  & 5 & 0.4 & 46 & 2.6$\times$10$^4$\\
3.1  & 10 & 0.5 & 55 & 2.2$\times$10$^4$\\
3.3  & 10 & 0.6 & 60 & 2.0$\times$10$^4$\\
\hline
\end{tabular}
  \label{tab:shape-functions}
\end{table}

Increasing of the groove depth up to 10~nm results in essential growth of the threshold power, up to 0.5-0.6~mW, while for the 5 nm depth it is close to the threshold of the microdisk without a groove defect (see Table~\ref{tab:threshold}). However, the modes with radial number $q = 2$ are not completely suppressed in this case.

Mode spacing of 15~nm is observed, when all the modes with $q > 1$ are suppressed. This spectral separation can be sufficient for applications of WGM microlasers. For example, arrays of self-organized InAs/InGaAs QDs have a temperature-induced shift of emission wavelength of about 0.45~nm/$\degree$C. Therefore, temperature increment by 30$\degree$C results in the red shift of the mode spectra by approximately 13~nm. Thus, the spacing of 15 nm is sufficient to prevent the mode forced by variation of the ambient temperature or device self-heating.

In conclusion, we have demonstrated that the mode spacing in a QD microdisk laser can be essentially increased by means of radial grooves etched by focused ion beam. Etching the groves results in the increase of lasing threshold and the decrease of fundamental mode quality factor by 1.5 and 2 times, respectively. Total suppression of all the modes except fundamental radial one can be achieved via a proper choice of geometrical parameters and the position of the groove defect.

\acknowledgments
The program of Fundamental Research in Nanotechnology and Nanomaterials of the Russian Academy of Science; RFBR (Project No. 14-02-01223, 15-32-20238, 13-02-12032); Federal Program on Support of Leading Scientific Schools (NSh-5062.2014.2).

\bibliography{references}

\end{document}